\documentclass[english,aps,floats,twocolumn,showpacs,nofootinbib]{revtex4}
\usepackage{pslatex}
\usepackage[T1]{fontenc}
\usepackage[latin1]{inputenc}
\usepackage{graphicx}
\usepackage{epsfig}

\usepackage{calc}
\usepackage{ifthen}

{
{
{
\newcommand{\bea}{\begin{eqnarray}}
\newcommand{\eea}{\end{eqnarray}}

\newcommand{\nc}{\newcommand}
\nc{\renc}{\renewcommand}
\nc{\eqs}[2]{\mbox{Eqs.~(\ref{#1},\,\ref{#2})}}
\nc{\eq}[1]{\mbox{Eq.~(\ref{#1})}}
\nc{\figs}[2]{\mbox{Figs.~(\ref{#1},\,\ref{#2})}}
\nc{\fig}[1]{\mbox{Fig~.(\ref{#1})}}
\nc{\be}[1]{\begin{equation} \mbox{$\label{#1}$}}
\nc{\ee}{\vspace{0.1cm}\end{equation}}

\newcommand{\bean}{\begin{eqnarray*}}
\newcommand{\eean}{\end{eqnarray*}}

\def\bfx{{\bf x}}
\def\bfk{{\bf k}}
\def\bfl{{\bf l}}
\def\ucl{u^{cl}}

\def\lae{\;^{<}_{\sim} \;}

\begin{document}
\title{Space-Dependent Step Features: Transient Breakdown of Slow-roll, Homogeneity and Isotropy During Inflation
}
\author{Rose Lerner}
\email{r.lerner@lancaster.ac.uk}
\author{John McDonald}
\email{j.mcdonald@lancaster.ac.uk}
\affiliation{Cosmology and Astroparticle Physics Group, University of
Lancaster, Lancaster LA1 4YB, UK}
\begin{abstract}

         A step feature in the inflaton potential can model a transient breakdown of slow-roll inflation.
Here we generalize the step feature to include space-dependence, allowing it also to model a breakdown of homogeneity and isotropy. The space-dependent inflaton potential generates a classical curvature perturbation mode characterized by the wavenumber of the step inhomogeneity. For inhomogeneities small compared with the horizon at the step, space-dependence has a small effect on the curvature perturbation. Therefore the smoothly oscillating quantum power spectrum predicted by the homogeneous step is robust with respect to sub-horizon space-dependence. For inhomogeneities equal to or greater than the horizon at the step, the space-dependent classical mode can dominate, producing a curvature perturbation in which modes of wavenumber determined by the step inhomogeneity are superimposed on the oscillating power spectrum. Generation of a space-dependent step feature may therefore provide a mechanism to introduce primordial anisotropy into the curvature perturbation.
Space-dependence also modifies the quantum fluctuations, in particular via resonance-like features coming from mode coupling to amplified superhorizon modes. However these effects are small relative to the classical modes.

\end{abstract}
\maketitle
\section{Introduction}

       Most models for the origin of the primordial density perturbation are based on a slowly-rolling, homogeneous inflaton field. However, in view of our incomplete understanding of physics at the energy scales of inflation, combined with the increasing precision of cosmic microwave background (CMB) observations, it is important to consider other possibilities for the origin of the primordial density perturbation. Various alternatives to the orthodox scenario have been considered, including features in the inflaton potential \cite{adams,step,step2,peiris,joy,features}, transient effects of a pre-inflationary era \cite{transient}, and trans-Planckian signatures \cite{tplanck}. In particular, a transient departure from slow-roll inflation has been considered, modelled by step feature in the inflaton potential \cite{adams,peiris,step,step2}. Here we will consider a more general possibility: a transient departure from homogeneity and isotropy in addition to slow-roll, modelled by an inflaton potential which is space-dependent during the step.

     The step feature considered in \cite{adams,step,step2,peiris} was originally motivated by the idea that a second scalar field $\psi$ could couple to the inflaton $\phi$ and undergo a phase transition. The phase transition is triggered by the inflaton field in a manner similar to the transition of the waterfall field of hybrid inflation models \cite{hi,hunt}. The $\psi$ field acquires a tachyonic mass and undergoes a phase transition towards the $\phi$-dependent minimum of its potential. The energy density as a function of $\phi$ will then decrease until $\psi$ reaches its $\phi$-independent minimum. The effect of this step down in energy density as a function of $\phi$ is modelled by a step feature in the single-field effective potential for $\phi$.
However, this reasoning also suggests that the step feature can have a space-dependence. Hybrid inflation-type phase transitions commonly proceed via tachyonic preheating \cite{tp}, where tachyonic amplification of sub-horizon quantum fluctuations of $\psi$
results in an inhomogeneous energy density \cite{tp,hitp}. In this case we expect the energy density
as a function of $\phi$ to be space-dependent during the step.
Therefore if a step feature can qualitatively model the effect of a phase transition on the inflaton, a space-dependent step feature is equally well-motivated.

    Regardless of any specific motivation, a space-dependent step feature can also be considered as a purely phenomenological model for a transient breakdown of homogeneity and isotropy. The issue then is how the space-dependence generically modifies the curvature perturbation relative to the homogeneous step.

  Our paper is organised as follows. In Section II we introduce and discuss the space-dependent step model.
In Section III we derive the equations for the classical and quantum modes. In Section IV we solve these equations for their effect on the curvature perturbation. In Section V we discuss our conclusions.

\section{A Space-Dependent Step Feature}

   In order to study how space-dependence alters a homogeneous step model, we will focus on the well-known $\phi^2$ chaotic inflation step of \cite{adams,step,step2}. The potential of this model is given by
\be{e1}   V_{step}(\phi) = \frac{1}{2} m^2 \phi^2 \left( 1+ c\, {\rm tanh} \left(\frac{\phi - b}{d}\right) \right)    ~,\ee
where $b$ is the value of $\phi$ at which the step occurs, $d$ parameterizes the duration of the step and $c$ parameterizes the height of the step. We generalise this by adding a space-dependent term which modulates the step,
\be{e2a} V(\phi) = V_{step}(\phi) + F(\phi)\sin(k_{L}x)\sin(k_{L}y)\sin(k_{L}z)     ~,\ee
where
\be{e3} F(\phi) = \frac{1}{2} g(\overline{\phi}) m^2 \phi^2  c\, {\rm tanh} \left(\frac{\phi - b}{d}\right)    ~\ee
The factor $g(\overline{\phi})$ parameterizes the magnitude and rate of dissipation of the inhomogeneities in the step,  where $\overline{\phi}(t)$ is the classical background field. This corresponds to a lattice of inhomogeneities of characteristic length scale $\lambda_{L} = 2 \pi/k_{L}$. Although approximately isotropic, a degree of anisotropy is inevitable once inhomogeneity is introduced. In practice, in order to simplify the numerical analysis, we will consider
an inhomogeneity which is planar in form, varying only in the $\bfk_{L}$ direction,
\be{e3a} V(\phi) = V_{step}(\phi) + F(\phi)\sin(\bfk_{L}.\bfx)     ~.\ee
In Appendix A we show that the mode equations for the planar inhomogeneity are similar to those of the approximately isotropic inhomogeneity of \eq{e2a}. The second term in \eq{e3a} effectively replaces $c$ with a space-dependent height for the step, $c(1 + g(\overline{\phi})\sin(\bfk_{L}.\bfx) )$. We will consider the case where the inhomogeneities exist over the duration of the step and the variation of the height of the step with $\bfx$ is approximately $c_{F}$, which can be modelled by setting  $g(\overline{\phi}) = (c_{F}/c)\exp\left(-\left(\overline{\phi} - b\right)^{2}/d^{2}\right)$, where we assume $c_{F} \leq c$. This rapidly tends to zero for $|\overline{\phi} - b| > d$. Note that we only use the background field in $g(\overline{\phi})$  since here $\phi$ serves only to parameterize the time-dependence of the inhomogeneities and is therefore not a quantum field.

\section{Mode Equations for the Classical and Quantum Modes}

     There will be two contributions to the curvature perturbation from the space-dependent step.
First, since the classical inflaton field is rolling in a space-dependent potential,
it will acquire a classical spatial mode of initial wavenumber equal to the inhomogeneity wavenumber $\bfk_{L}$.
Second, there will be quantum fluctuations of the inflaton amplified by the step as in the homogeneous case, but with additional features due to the space-dependence of the step.

      To study the combined metric and field perturbations, we introduce the Mukhanov
variable $u = -z {\cal R}$, where $z = a \dot{\phi}/H$ and ${\cal R}$ is the curvature perturbation \cite{muk}.
The space-dependence of the inflaton is defined to be entirely in the perturbation $u(\bfx,\tau)$ about the homogeneous background. The action for $u$ in the absence of the space-dependent potential is then \cite{muk}
\be{e4}  S = \frac{1}{2} \int \left( \partial_{\mu} u \partial^{\mu} u + \frac{z^{''}}{z} u^2  \right) d^{4}x  ~,\ee
where indices are with respect to conformal time $\tau$ and comoving coordinates $\bfx$, and
\be{e5} \frac{z^{''}}{z} = -\frac{2 H^{'}}{H} \frac{\phi^{''}}{\phi^{'}} - 2 a H^{'} + 2 a^{2} H^{2} + \frac{2 H^{'\;2}}{H^2} - \frac{H^{''}}{H} - a^2 \frac{d^{2}V_{step}}{d\phi^{2}}    ~.\ee
In this primes denote derivatives with respect to $\tau$ and $\phi = \phi(t)$. Note that $\overline{\phi}$ is treated as constant in the derivatives of $V_{step}$ but is subsequently set equal to $\phi(t)$. \eq{e4} is of the same form as for a real scalar field in flat space with a time-dependent mass squared $ m_{u}^{2} = - z^{''}/z$.
When the space-dependence is included, the effect is to modify the last term
\be{e6} \frac{z^{''}}{z}u^2 \rightarrow  \frac{z^{''}}{z}u^2
 - a^{2} \frac{d^{2}F}{d \phi^{2}} {\rm sin(\bfk_{L}.\bfx)}u^2 - 2 a^{3} \frac{dF}{d \phi} {\rm sin(\bfk_{L}.\bfx)}u
 ~.\ee
We then derive the equations for the modes including space-dependence.
Expanding $u(\bfx, \tau)$ in classical modes
\be{e11}   u(\bfx,\tau) = \frac{1}{\left(2 \pi\right)^{3/2}}
\int d^{3}k\, u_{\bfk}\!(\tau)\,e^{i \bfk.\bfx}     ~\ee
and substituting into the equation for $u(\bfx,\tau)$ with the space-dependence included,
\be{e12}
u^{''} - \nabla^{2}u = \frac{z^{''}}{z}u - a^{2} \frac{d^{2}F}{d \phi^{2}} {\rm sin(\bfk_{L}.\bfx)} u
- a^{3} \frac{dF}{d \phi} {\rm sin(\bfk_{L}.\bfx)}
  ~,\ee
we obtain a system of coupled equations for the classical mode functions
$$ u_{\bfk}^{''} + \bfk^{2} u_{\bfk} - \frac{z^{''}}{z} u_{\bfk}
+ \frac{a^{2}}{2 i} \frac{d^{2}F}{d \phi^{2}}
  \left(  u_{\bfk - \bfk_{L}} - u_{\bfk + \bfk_{L}} \right)
$$
\be{e13}
+ \frac{a^{3}}{2 i} \frac{dF}{d \phi}
\left(2\pi\right)^{3/2} \left(  \delta \left(\bfk -\bfk_{L}\right) - \delta \left(\bfk + \bfk_{L}\right) \right)
= 0   ~.\ee
The last term, which applies only to the modes with $\bfk = \pm \bfk_{L}$, is due to the rolling of the inflaton in the space-dependent potential, which creates a classical inflaton mode with $\bfk = \pm \bfk_{L}$.

 To quantize the modes, we will consider the action $S$ without the linear term in $u$. The linear term only affects modes which are integer multiples of $\bfk_{L}$. Therefore we can consider the
modes with $\bfk \neq n\bfk_{L}$ to be entirely quantum in origin and due to the terms in the action which are quadratic in $u$, while the modes with
$\bfk = n\bfk_{L}$ can be either quantum or classical, depending on which contribution is dominant.

At times well before and after the step when the space dependence is effectively zero, the field may be expanded in terms of creation and annihilation operators with respect to the vacuum at an initial time, defined as $N_{i}$ e-foldings before the end of inflation,
\be{e7}  u(\bfx,\tau) = \int \frac{d^{3}k}{\left(2 \pi\right)^{3/2}}
\left[ u_{\bfk}(\tau) e^{i\bfk.\bfx} a_{\bfk} + u_{\bfk}^{*}(\tau) e^{-i\bfk.\bfx} a_{\bfk}^{\dagger} \right]    ~.\ee
The field operator $u(\bfx,\tau)$ and its canonical conjugate $\pi(\bfx,\tau) = u^{'}(\bfx,\tau)$ satisfy the usual equal-time commutation relations and Hamiltonian field equations.

The quantum modes for $\bfk \neq n \bfk_{L}$ satisfy \eq{e13} without the last term.
\be{e13a} u_{\bfk}^{''} + \bfk^{2} u_{\bfk} - \frac{z^{''}}{z} u_{\bfk}
+ \frac{a^{2}}{2 i} \frac{d^{2}F}{d \phi^{2}}
  \left(  u_{\bfk - \bfk_{L}} - u_{\bfk + \bfk_{L}} \right)
= 0   ~.\ee
It is not immediately obvious that the quantum mode functions will satisfy the classical field equations. In the case without space-dependence, when substituting \eq{e7} into the classical field equation, the creation and annihilation operators factor out. Including the space dependence introduces mode mixing and means this is no longer true. However, by performing a unitary transformation to new canonical coordinates and conjugate momenta, it is shown in Appendix B that the late-time solutions for the quantum modes $u_{\bfk}$ are indeed the solutions of the classical mode equations \eq{e13a} obtained using the quantum mode initial conditions. The initial conditions are obtained assuming a massless field with the modes of interest initially well within the horizon ($\bfk^{2} \gg a^2 H^2$), giving
\be{e10}   u_{\bfk} = \frac{e^{-i \left(|\bfk|\tau + \theta_{\bfk}\right)}}{\sqrt{2 |\bfk|}}       \ee where $\theta_{\bfk}$ is a random phase.
There will also be quantum contributions to the modes at $\bfk = n\bfk_{L}$. In this case we assume that the larger of the classical and quantum contributions dominates.

We have included a phase $\theta_{\bfk}$ for each mode. In the case without space-dependence this phase has no physical effect and is usually set to zero. However, as we have shown, space-dependence implies that different modes couple, in which case the phases will affect the solutions for $u_{\bfk}(\tau)$. In absence of phases, the quantum mode solutions $u_{\bfk}$ would generate an ensemble of
classical universes with the classical field mode probabilities determined from a Gaussian distribution with mean value $|u_{\bfk}|$ \cite{pol}.
We would then interpret the observed Universe as one of the ensemble of possible classical universes.
When we include the initial phases and mode mixing at the step, each set $i$ of initial random phases will generate
a distinct solution for the modes at late times, $u_{\bfk,\;i}$. Each of these solutions should be thought of as producing a distinct set of classical states which contributes to the total ensemble of classical states. Measurement of the state of the universe at late times will then select out a classical universe related to the solution $u_{\bfk,\;i}$ for one particular set of initial phases. Therefore in the following we will compute results for one set of random initial phases $\theta_{\bfk}$.

\section{Numerical Solution of the Mode Equations}
\label{4}

\section*{(i) Classical Modes}

      The space-dependence of the step potential will imprint a classical space-dependence on the inflaton of
wavenumber $\bfk_{L}$. The coupling of the modes in \eq{e13} will then induce further classical modes of
wavenumber $n\bfk_{L}$. In order to analyse these modes, we will expand the classical field as
\be{cl1}  u(\bfx,\tau) =   \sum_{n = -\infty}^{\infty} u_{n}^{cl}(\tau)e^{i n\bfk_{L}.\bfx}      ~,\ee
which gives, on substituting into \eq{e12},
$$ u_{n}^{cl\;''} + (n \bfk_{L})^{2} \ucl_{n} - \frac{z^{''}}{z} \ucl_{n}
+ \frac{a^{2}}{2 i} \frac{d^{2}F}{d \phi^{2}}
  \left(  \ucl_{n-1} - \ucl_{n+1} \right)
$$
\be{cl2}
+ \frac{a^{3}}{2 i} \frac{dF}{d \phi}
\left(  \delta_{n,1}  - \delta_{n,-1} \right)
= 0   ~.\ee
Note that $u_{n}^{cl}(\tau)$ shares the same dimensions with $u(\bfx,\tau)$ but not with the quantum modes $u_{\bfk}(\tau)$.
We will consider throughout the case of a mode $\bfk = (k,0,0)$ in the $x$-direction, with $\bfk_{L}= (k_{L},0,0)$, where $k$ and $k_{L}$ are positive.

          We have solved these equations numerically for the case of $c = c_{F} = 0.002$, such that the spatial fluctuation of the step potential is of the order of the mean potential. (The maximum value of $c$ consistent with observations in the case of the homogeneous step is $0.003$ \cite{step}.)

\begin{figure}[htb]
                    \centering
                    \includegraphics[width=0.45\textwidth, angle=0]{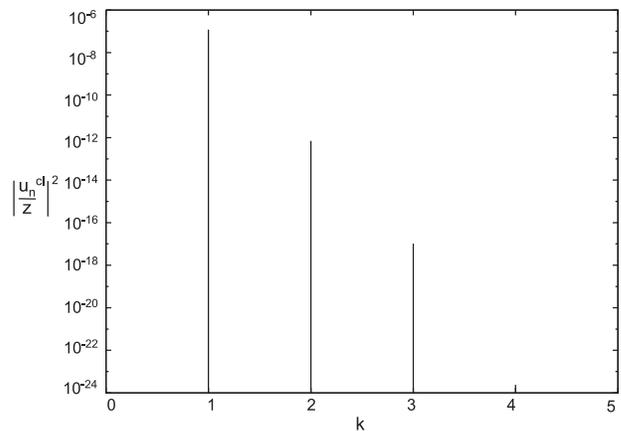}
                    \caption{\footnotesize{$|u_n^{cl}/z|^2$ as a function of $k \equiv n k_{L}$ for $c = c_{F} = 0.002$ and $k_L = 1$, where $k$ is in units of $aH|_{STEP}$. This corresponds to a step with inhomogeneities of wavelength equal to the horizon at the step.}
                    \label{figN1}}
\end{figure}
    In Figure~\ref{figN1} we show $|\ucl_{n}/z|^2$ for the case of a step with inhomogeneity wavelength equal to the horizon at the step, $k_{L} = \left.  a H \right|_{STEP}$. (We plot $k$ in units of $\left. a H \right|_{STEP}$.) A series of classical modes are generated with wavenumber $nk_{L}$. However, the secondary modes at $n=2,3,...$ are very much smaller than the $n = 1$ mode, which will dominate the curvature perturbation. The value of
$|\ucl_{n}/z|$ for this mode is slightly larger that the conservative
limit following from the observed value of $<{\cal R}^2>$, $|\ucl_{n}/z| < 10^{-4}$ (the relationship between the modes and the curvature perturbation is discussed in Appendix C), and therefore makes a larger contribution to $<{\cal R}^2>$ than the homogeneous step. In this case the features in the curvature perturbation power spectrum will be quite different from the homogeneous step, with domination by a single perturbation mode of wavenumber $\bfk_{L}$ rather than the oscillations expected for the homogeneous step.

       For smaller values of $c_{F}$ we find that the height of $|\ucl_{n}/z|^2$ decreases as $c_{F}^2$. Therefore if the spatial fluctuation of
the step potential is significantly smaller than the homogeneous step in the potential, the effect of the classical modes will be negligible.  However, for $c_{F}$ not too much smaller than $c$, the curvature perturbation will be a combination of an oscillating power spectrum of quantum origin plus a single classical mode at $\bfk = \bfk_{L}$.
\begin{figure}[htb]
                    \centering
                    \includegraphics[width=0.45\textwidth, angle=0]{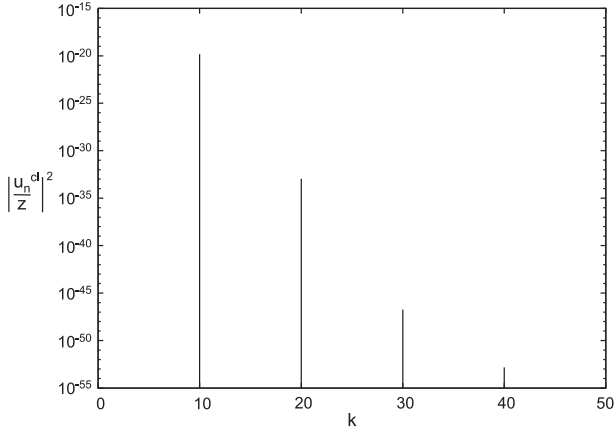}
                    \caption{\footnotesize{$|u_n^{cl}/z|^2$ as a function of $k \equiv n k_{L}$ for $c = c_{F} = 0.002$ and $k_L = 10$. This corresponds to a step with inhomogeneities of wavelength 10 times less than the horizon at the step.}
                    \label{figN2}}
\end{figure}

\begin{figure}[htb]
                    \centering
                    \includegraphics[width=0.45\textwidth, angle=0]{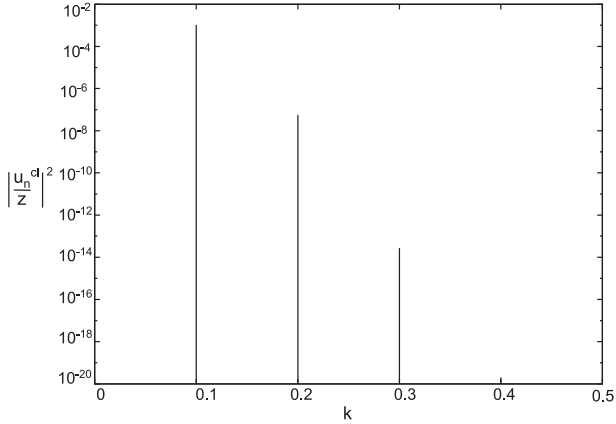}
                    \caption{\footnotesize{$|u_n^{cl}/z|^2$ as a function of $k \equiv n k_{L}$ for $c = c_{F} = 0.002$ and $k_L = 0.1$. This corresponds to a step with inhomogeneities of wavelength 10 times greater than the horizon at the step.}
                    \label{figN3}}
\end{figure}
       In Figures~\ref{figN2} and \ref{figN3} we show the effect of
varying the length scale of the inhomogeneities relative to the horizon at the step. Figure~\ref{figN2} shows $|\ucl_{n}/z|^2$ for an inhomogeneity of wavelength equal to 0.1 times the horizon at the step. There is a dramatic reduction in $|\ucl_{n}/z|$ by a factor of $10^{-6}$ compared with the horizon-sized step. Since in the case of steps motivated by phase transitions we expect the inhomogeneities to be due to tachyonic amplification of sub-horizon quantum fluctuations, this suggests that for such steps the
effect of space-dependence on the power spectrum will be negligible.
This is an important conclusion, as it means that the
oscillation features observed in \cite{step} are a robust prediction of phase transitions during inflation, even if they are of the more complex tachyonic preheating type. In Figure~\ref{figN3} we consider the possibility that the space-dependence during the step is larger than the horizon. In this case we find a strong amplification of $|\ucl_{n}/z|$ relative to the horizon-sized step.

    We conclude that the contribution of a classical spatial mode to the curvature perturbation is strongly dependent on its size relative to the horizon at the step. A significant contribution to the curvature perturbation is possible only if the size of the inhomogeneities is close to or larger than the horizon at the step. In this case we can expect to see a significant deviation from the smoothly oscillating power spectrum predicted by the homogeneous step feature. In particular, we expect the power spectrum will have a strong contribution from modes with wavenumber characterized by the inhomogeneity in the step.
For the case of a less idealized step feature which is approximately isotropic and can be
expanded as a range of modes with $|\bfk| \approx k_{L}$, the curvature perturbation will be dominated by a range of modes with $|\bfk| \approx k_{L}$ rather than a single mode.

\section*{(ii) Quantum Modes}

    We next consider the purely quantum modes due to the quadratic terms in the action for $u$. It will be convenient to change variable to $\alpha = \log(a)$, in which case \eq{e13a} becomes
$$ u_{\bfk \; \alpha \alpha} + \left(\frac{H_{\alpha}}{H} + 1 \right) u_{\bfk\; \alpha}  +
\left[ \frac{\bfk^{2}}{e^{2 \alpha}H^{2}} - \left[2 - \frac{4 H_{\alpha}}{H} \frac{\phi_{\alpha \alpha}}{\phi_{\alpha}} - 2 \left( \frac{H_{\alpha}}{H}\right)^{2} \right. \right.$$
\be{e14} \left. \left. - \frac{5 H_{\alpha}}{H}  - \frac{1}{H^{2}} \frac{d^{2}V_{step}}{d\phi^{2}} \right]     \right]u_{\bfk} + \frac{1}{H^{2}} \frac{d^{2} F}{d \phi^{2}} \frac{1}{2 i} \left(u_{\bfk - \bfk_{L}}
 - u_{\bfk + \bfk_{L}} \right)
= 0 ~.\ee
$\phi(t)$ evolves according to
\be{e15}  \phi_{\alpha \alpha} +  \left(\frac{H_{\alpha}}{H} + 3 \right) \phi_{\alpha} +
\frac{1}{H^{2}} \frac{d V_{step}}{d\phi}(\phi)  =  0    ~,\ee
\be{e16} H_{\alpha} = - 4 \pi G \phi_{\alpha}^{2}    ~.\ee
The subscript $\alpha$ denotes differentiation with respect to $\alpha$.
We have solved the system of equations \eq{e14}, \eq{e15} and \eq{e16} for a range of values of $k$ of interest. To do this we have used a system of seven equations for each $\bfk$, corresponding to full equations for $u_{\bfk}$, $u_{\bfk \pm \bfk_{L}}$ and $u_{\bfk \pm 2 \bfk_{L}}$, plus the equations for $u_{\bfk \pm 3 \bfk_{L}}$ with the space dependence set to zero. Random values of the phase $\theta_{\bfk}$ were generated in order to study the effect of initial mode phases on the power spectrum.

   The initial conditions for the real and imaginary parts of $u_{\bfk}$ at $\bfk^{2} \gg a^{2} H^{2}$ follow from \eq{e10} and are
\be{e17}  u_{\bfk}^{1} = \frac{\cos(\theta_{\bfk} + |\bfk|\tau)}{\sqrt{2|\bfk|}} \;\;\;,\;\;\; u_{\bfk\;\alpha}^{1} =  \left. - \sqrt{\frac{|\bfk|}{2}}\frac{1}{e^{\alpha} H}\right. \sin(\theta_{\bfk} + |\bfk|\tau) ~,\ee
\be{e18}  u_{\bfk}^{2} = - \frac{\sin(\theta_{\bfk} + |\bfk|\tau)}{\sqrt{2|\bfk|}}  \;\;\;,\;\;\; u_{\bfk\;\alpha}^{2} = \left.  \sqrt{\frac{|\bfk|}{2}}\frac{1}{e^{\alpha} H}\right. \cos(\theta_{\bfk} + |\bfk|\tau) ~,\ee
where $u_\bfk = u_\bfk^1 + i u_\bfk^2$.

We solved for $u_\bfk$, and used this to calculate $|u_{\bfk}/z|^2$ and the primordial curvature perturbation power spectrum,
\be{power} {\cal P_R}^{1/2}(\bfk) = \sqrt{\frac{\left|\bfk\right|^3}{2\pi^2}}\left|\frac{u_\bfk}{z}\right| ~.\ee
Although the power spectrum in the case of an anisotropic perturbation does not have a direct interpretation in terms of the mean squared curvature perturbation per logarithmic interval in $k$, we have plotted ${\cal P_{R}}(k)$ with $k \equiv k_{x}$ in order to compare with the results for the homogeneous step \cite{step}. As discussed in Appendix C, the more useful quantity for calculating the observed curvature perturbation is $|u_{\bfk}/z|^2$, which we also plot. After the step has completed, the superhorizon modes grow proportional to $z$, so $|u_\bfk/z|^2$ is constant. It is in this limit that we calculate ${\cal P_R}(k)$ and $|u_\bfk/z|^2$ for the superhorizon modes.

The two components of $u_\bfk$ oscillate prior to leaving the horizon, with a frequency that increases as $a^{-1}$ as we go to earlier times. The background field $\phi$ is initially slow-rolling and each mode is evolved from $|\bfk| >> aH$ to $|\bfk| << aH$, The initial time for each mode is chosen differently in order to avoid the prohibitively small step size required for high frequency oscillations at very early times. A second-order Runge-Kutta algorithm was found to provide accurate solutions of the system of equations.

\begin{figure}[htb]
                    \centering
                    \includegraphics[width=0.45\textwidth, angle=0]{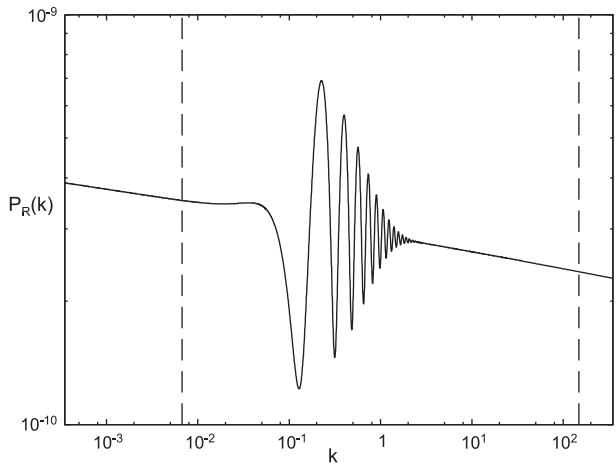}
                    \caption{\footnotesize{The curvature perturbation power spectrum for $c = 0.002$ and $c_F = 0$, where $k$ is in units of $aH|_{STEP}$. Scales accessible to observations are between the dashed lines.}
                    \label{figE}}
\end{figure}

Figure~\ref{figE} shows the results for the case without space-dependence, $c_F = 0$, which reproduces the results of \cite{adams}. In this and other figures we use $c=0.002$, $d=0.01$ and $m=2.5\times 10^{-6}$ (in units with  $~\hbar = c = G = 1$), where the value of $m$ is determined by the COBE normalized value of ${\cal P_R}$. We consider the observable range of $k$ to correspond to modes which leave the horizon from $N = 60$ to $N = 50$, indicated by the dashed lines, with the step centered at $N = 55$. The wavenumber which exits the horizon at the step occurs at the mid-point, $k = 1$ in units of $aH|_{STEP}$.

\begin{figure}[h]
                    \centering
                    \includegraphics[width=0.45\textwidth, angle=0]{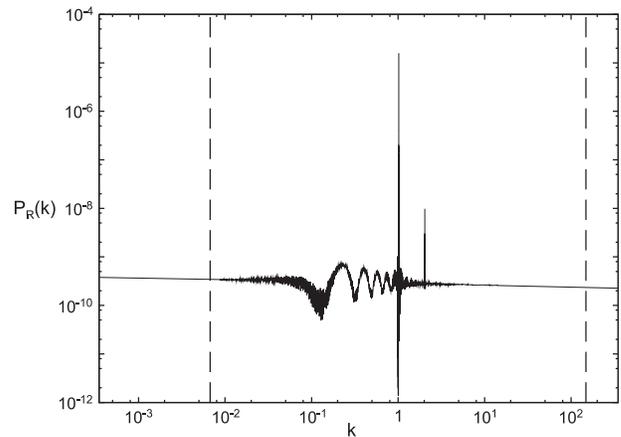}
                    \caption{\footnotesize{The curvature perturbation power spectrum for inhomogeneities of wavelength equal to the horizon at the step, $k_{L} = 1$, for the case $c_F = c = 0.002$.  Resonances are observed at $\bfk = \bfk_{L}$ and $\bfk = 2 \bfk_{L}$. The apparent broadening of the power spectrum is due to the random initial phase for each mode combined with mode coupling.}
                    \label{figA1}}
\end{figure}

\begin{figure}[h]
                    \centering
                    \includegraphics[width=0.45\textwidth, angle=0]{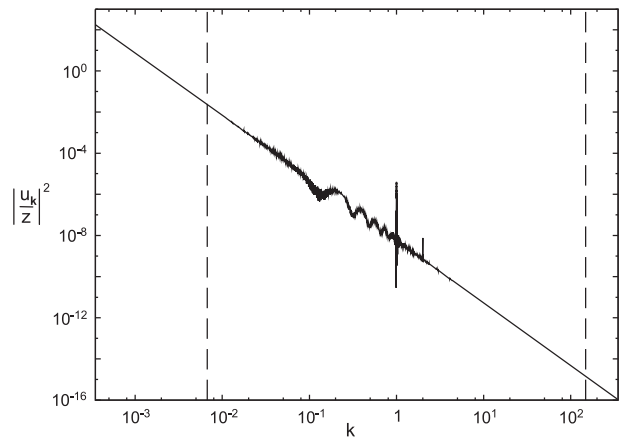}
                    \caption{\footnotesize{$|u_{\bfk}/z|^2$ for inhomogeneities of wavelength equal to the horizon at the step, $k_{L} = 1$, for the case $c_F = c = 0.002$.}
                    \label{figA2}}
\end{figure}

Figure~\ref{figA1} shows the curvature perturbation power spectrum ${\cal P}_{{\cal R}}(k)$ with $k = k_{x}$, for the case of a space-dependent step of wavelength equal to the horizon at the step ($k_L = 1$).
In Figure~\ref{figA2} we show the corresponding plot of $|u_{\bfk}/z|^2$.
There are two notable new features in the power spectrum. In addition to the oscillation feature of Figure~\ref{figE}, we observe (a) sharp resonance-like features and (b) an apparent broadening of the spectrum relative to the line observed in the homogeneous case.

\noindent {\bf (a) Resonances:} The resonance-like features are found to occur at $k=k_L$, $k=2 k_L$ and $k = 3 k_L$. We expect that had we used modes up to $u_{\bfk+n\bfk_L}$ in the calculation of $u_\bfk$, we would observe $n$ resonances, decreasing in magnitude as $n$ increases. While having a large amplitude, the width of the resonances in $k$ is very narrow, which greatly diminishes their effect on the curvature perturbation.
\begin{figure}[h]
                    \centering
                    \includegraphics[width=0.45\textwidth, angle=0]{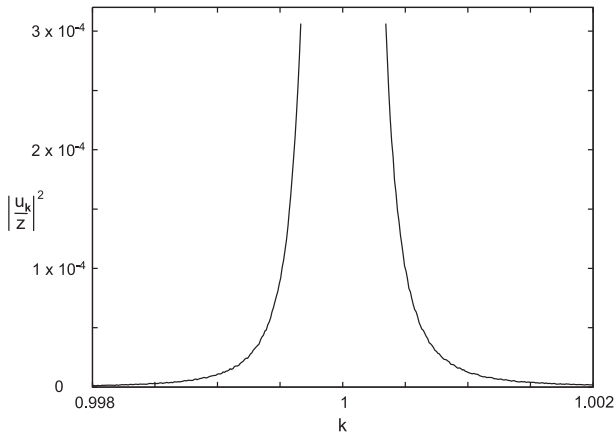}
                    \caption{\footnotesize{A close-up of the $\bfk = \bfk_L$ resonance of Figure~\ref{figA2}.}
                    \label{figD}}
\end{figure}

The origin of the resonance is the coupling of a mode $u_{\bfk}$ to a much larger superhorizon mode $u_{\bfk-\bfk_{L}}$, where $|\bfk - \bfk_{L}| \ll aH$. Since the superhorizon growing mode has been strongly amplified, with
$u_{\bfk-n\bfk_L} \propto z \propto a$ since exiting the horizon, it provides a strong driving term in the $u_{\bfk}$ mode equation, producing rapid growth of the mode. The longer the wavelength of the  $\bfk - \bfk_{L}$ mode, the larger will be the amplification of  $u_{\bfk-\bfk_{L}}$. Therefore the $u_{\bfk}$ resonance will tend to large values as $|\bfk - \bfk_{L}| \rightarrow 0$.  A close-up of the $k = k_L$ resonance of Figure~\ref{figA1} is shown in Figure~\ref{figD}. (A lower bound $5 \times 10^{-4}$ has been imposed on $|\bfk -\bfk_L|$ when plotting the figures.) We find from Figure~\ref{figD} that $|u_{\bfk}/z|^2 \propto |\bfk-\bfk_L|^{-3}$. The resonance depends only on $|\bfk-\bfk_L|$ because the amplified superhorizon mode driving the growth of the resonance depends only on $|\bfk-\bfk_L|$.

   Although resonance due to coupling to large superhorizon modes is a theoretically interesting consequence of the space-dependent step, we believe it will make a negligible contribution to the curvature perturbation compared to the classical modes with $\bfk = n\bfk_{L}$. In Appendix C it is shown that a quantum resonance produces an effective classical mode given by
\be{e18a} |\ucl_{n\;eff}|^2 = \frac{1}{\left(2 \pi\right)^3}
\int_{0}^{k_{L}} |u_{\bfk}|^2 4 \pi k^{'\;2} d k^{'}  \;\; ; \;\;  k^{'} \equiv |\bfk - n \bfk_{L}| ~.\ee
The dependence of $|u_{\bfk}/z|^2$ on $k^{'}$ is critical. If we integrate \eq{e18a} from a lower bound $k^{'} = \delta k$ to $k_{L} \gg \delta k$, then with $|u_{\bfk}/z|^2 = A_{s}/k^{' s}$ we obtain
$$ \left|\frac{\ucl_{n\;eff}}{z}\right|^2 = \frac{A_{s}}{2 \pi^2} \frac{\delta k^{3-s}}{s-3} \;\;; \;\; s > 3   ~,$$
\be{e18b} \left|\frac{\ucl_{n\;eff}}{z}\right|^2 = \frac{A_{3}}{2 \pi^2} \ln \left( \frac{k_{L}}{\delta k} \right) \;\; ; \;\; s = 3 ~.\ee
Therefore if $s = 3$, as our numerical results indicate, the logarithmic dependence on the cut-off will
prevent a large enhancement as $\delta k \rightarrow 0$. We generally expect a physical lower bound on $\delta k$ to exist, corresponding to the
largest unsuppressed superhorizon mode. For example, the amplitude of superhorizon modes in $\phi^2$ chaotic inflation will become suppressed by effect of the finite mass term, while more generally the largest amplified mode will correspond to the mode exiting the horizon at the onset of inflation. In this case the logarithm in
\eq{e18b} can only enhance the prefactor by a few orders of magnitude at most. If $s$ were significantly larger than 3 then a large enhancement might be possible as $\delta k \rightarrow 0$. However, we find no evidence for
$s > 3$ from our numerical results.

   From Figure~\ref{figD}, assuming that $|u_{\bfk}/z|^2 \propto k^{'\;3}$ for all $k^{'} \geq \delta k$,  we find $A_{3} = 4 \times 10^{-14}$ and so for $n = 1$
\be{e18d}     \left|\frac{\ucl_{1\;eff}}{z}\right|^2 = 2 \times 10^{-15}  \ln \left( \frac{k_{L}}{\delta k} \right)   ~.\ee
For any reasonable value of $\delta k$ this will be much smaller than the $n =1$ classical mode and also much smaller than the observational upper bound of $10^{-8}$.

\begin{figure}[h]
                    \centering
                    \includegraphics[width=0.45\textwidth, angle=0]{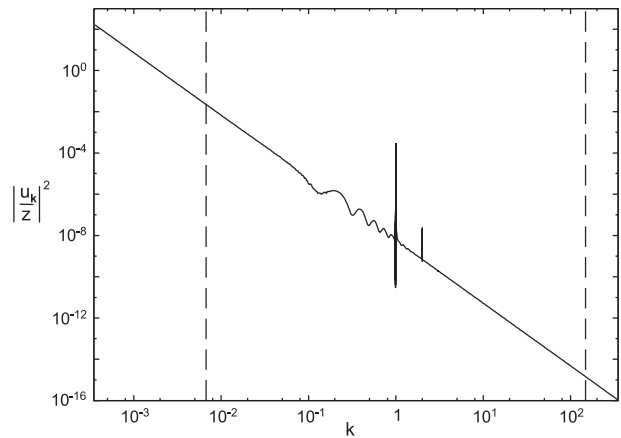}
                    \caption{\footnotesize{$|u_{\bfk}/z|^2$ averaged over many sets of random phases for inhomogeneities of wavelength equal to the horizon at the step, $k_{L} = 1$, for the case $c_F = c = 0.002$. The apparent broadening in Figure~\ref{figA2} is no longer seen.}
                    \label{figA3}}
\end{figure}
\begin{figure}[htb]
                    \centering
                    \includegraphics[width=0.45\textwidth, angle=0]{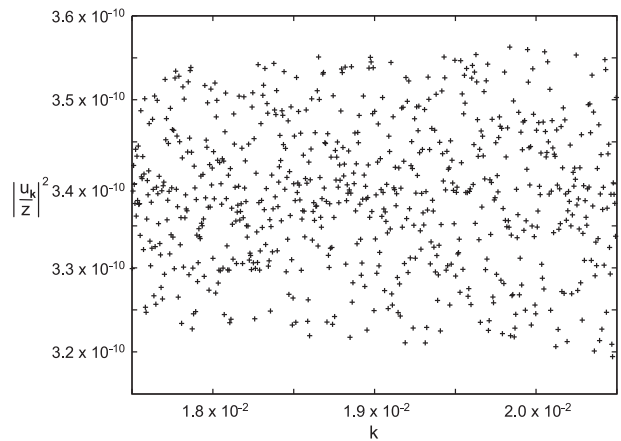}
                    \caption{\footnotesize{Close-up of Figure~\ref{figA2}, showing the discrete random fluctuation of the power spectrum with each step in k.}
                    \label{figF}}
\end{figure}
\noindent {\bf (b) Broadening:}  The apparent broadening of the power spectrum is due to the random initial phases of the modes. For each value of $\bfk$, the random phase produces a value of ${\cal P_R}^{1/2}(\bfk)$ which is shifted either above or below the homogeneous result. In Figure~\ref{figA3} we show the same plot as in Figure~\ref{figA1} but averaged over many sets of initial random phases. Seen close up, the power spectrum is actually a series of discrete points, randomly above or below the mean spectrum, at values of $k$ separated by the step size of the numerical solution.
This can be seen in Figure~\ref{figF} (a close up of Figure~\ref{figA2} for $1.75 \times 10^{-2} \leq k \leq 2.05 \times 10^{-2}$). There is no apparent correlation between ${\cal P_R}$ for different values of $k$ - it is a random scatter about the result for zero phase. This scatter is due to mode coupling, with each mode having a completely random initial phase. In the case of a homogeneous step there is no mode coupling, therefore the initial phases can be factored out of the mode equations and have no observable effect. We have produced solutions using several different sets of random initial phases and find no qualitative difference in the resulting power spectrum, although the detailed pattern of random fluctuations depends on the initial phases.

In practice, when translating the primordial power spectrum into physical observables, we average over the contributions of modes for a finite range of $k$. Since the random fluctuations will average out to zero over this finite range of $k$, it is likely that the observable effects will be indistinguishable from the case where all initial phases are set to zero (which is equivalent to averaging over many sets of initial random phases).
However, the transient coupling of modes during the step may also lead to some correlation of the phases of the modes. This could generate non-Gaussanity in the curvature perturbation.

\section{Discussion and Conclusions}
\label{5}

   In this paper we have been concerned with how space-dependence of the inflaton potential can manifest itself in the curvature perturbation.

     A space-dependent step feature provides a model for a transient breakdown of slow-roll, homogeneity and isotropy during inflation. We have studied how the curvature perturbation of the well-known $\phi^2$ chaotic inflation step feature of \cite{step} is modified by space-dependence.

    We find that the effect of space-dependence depends strongly on the size of the inhomogeneity in the step potential relative to the horizon at the step. The curvature perturbation can be dominated by the effect of the space-dependence if the inhomogeneities are close to or larger than the horizon, but their effect is strongly suppressed if the inhomogeneities are small compared with the horizon.

     One motivation for step features is a phase transition occurring during inflation. If the phase transition was of the tachyonic preheating type, we would expect the step in the inflaton
potential to be strongly space-dependent. However, the space-dependence would typically be on sub-horizon scales, coming from tachyonic amplification of sub-horizon quantum fluctuations. If the space-dependence is much smaller than the horizon, our results imply that space-dependence will have a negligible effect on the curvature perturbation power spectrum. Therefore the smoothly oscillating power spectrum expected for the homogeneous step is likely to be a robust prediction, typically unaffected by space-dependence during the phase transition.

      We also considered the possibility that the space-dependence is on scales equal to or greater than the horizon at the step. In this case the
curvature perturbation can be dominated by the effect of space-dependence. For the simple model of an anisotropic step
with inhomogeneities of wavenumber $\bfk_{L}$, the curvature perturbation will have a single classical
mode of wavenumber $\bfk_{L}$ superimposed on the oscillating power spectrum from the amplification of quantum fluctuations during the step.  As a result, the signature of space-dependent step features can be quite different from the smoothly oscillating power spectrum expected for a homogeneous step, with the curvature perturbation dominated by modes characterized by the wavenumber of the inhomogeneity.

The quantum modes are also modified by the effect of space-dependence and mode coupling. In particular, the quantum curvature perturbation has resonance-like features at integer multiples of the inhomogeneity wavenumber $\bfk_{L}$.
These are due to coupling to strongly amplified superhorizon modes.
Although the resonances appear large in a plot of the curvature perturbation power spectrum, they make only a very small contribution to the curvature perturbation compared with that from the classical modes, as a result of their narrow width. A second effect of mode coupling during the step is that the
curvature perturbation becomes dependent upon the initial phases of the quantum modes, producing a scattering (`broadening') of the curvature perturbation modes as a function of $\bfk$. However, this random scattering is likely to average to zero in any physical observable.

       There may be other effects of a space-dependent step. One is that the transient coupling of modes during the step could induce some correlation between the phases of the modes, generating a degree of non-Gaussianity in the
curvature perturbation. It has been shown that non-Gaussianity is  generally produced by step features in single-field inflation models
 \cite{stepNG1,stepNG2} even in the absence of space-dependence.
In addition, an oscillating potential may produce non-Gaussianity in sub-horizon modes at the step \cite{stepNG2}. It will therefore be important to establish if non-Gaussanity from space-dependence can be distinguished from these other sources. A second effect is that an anisotropic space-dependent step feature could provide a mechanism to generate primordial anisotropy in the curvature perturbation. It has recently been claimed that there is a large scale statistical anisotropy in the observed CMB \cite{asym}. One explanation of this is the existence of a large superhorizon mode in the curvature perturbation along a particular direction $\bfk$ \cite{kam}. This would require a mechanism to break isotropy and to amplify a single mode. Generation of an anisotropic space-dependent step of wavenumber $\bfk_{L}$ would be one way to introduce such a feature into the curvature perturbation, with modes with $\bfk \approx \bfk_{L}$ being generated by the step. Although an inflaton field cannot account for the observed anisotropy without introducing too much CMB inhomogeneity, a slowly rolling curvaton field with a space-dependent step in its potential might be able to \cite{kam} \footnote{The possibility of a curvaton with transient breakdown of slow-roll has recently been explored in \cite{matsuda}.}. A method of checking the presence of large scale anomalies in the observed CMB temperature has been suggested in \cite{poltest}. The polarization-temperature correlations presented there could be a way to test the step feature presented in this paper.

\section*{Acknowledgements}

This work was supported by the European Union through the Marie Curie Research and Training Network "UniverseNet" (MRTN-CT-2006-035863). The work of RL was also supported by an STFC studentship.

\renewcommand{\theequation}{A-\arabic{equation}}
 \setcounter{equation}{0}

\section*{Appendix A: Mode equations for an approximately isotropic distribution of inhomogeneities}

    In our analysis we have considered mode equations for an anisotropic planar inhomogeneity in the step potential. In this appendix we generalise the mode equations to the case of an approximately isotropic distribution of inhomogeneities.
In this case the step potential is given by
\be{aa1} V(\phi) = V_{step}(\phi) + F(\phi)\sin(k_{L}x)\sin(k_{L}y)\sin(k_{L}z)     ~.\ee
This corresponds to a 3-D `lattice' of inhomogeneities of diameter $2 \pi/\lambda_{L}$. Strictly speaking this also violates isotropy, but far more weakly than
the planar inhomogeneity.
Following the same steps as before, the classical mode equations are now given by
$$  u_{\bfk}^{''} + \bfk^{2} u_{\bfk} - \frac{z^{''}}{z} u_{\bfk} + \frac{a^{2}}{(2 i)^{3}} \frac{d^{2}F}{d \phi^{2}}
\left[  \left(u_{\bfk - \bfk_{L}^{0}} - u_{\bfk + \bfk_{L}^{0}}\right) \right.$$
$$ \left. +  \left(u_{\bfk - \bfk_{L}^{x}} - u_{\bfk + \bfk_{L}^{x}}\right) + \left(u_{\bfk - \bfk_{L}^{y}} - u_{\bfk + \bfk_{L}^{y}}\right)
+ \left(u_{\bfk - \bfk_{L}^{z}} - u_{\bfk + \bfk_{L}^{z}}\right)
\right] $$
$$ + \frac{a^{3}}{(2 i)^{3}} \frac{dF}{d \phi}
\left[  \left(\delta_{\bfk, \bfk_{L}^{0}} - \delta_{\bfk, -\bfk_{L}^{0}}\right)
 +  \left(\delta_{\bfk,\bfk_{L}^{x}} - \delta_{\bfk,-\bfk_{L}^{x}}\right) \right. $$
\be{aa2} \left. + \left(\delta_{\bfk, \bfk_{L}^{y}} - \delta_{\bfk,-\bfk_{L}^{y}}\right)
+ \left(\delta_{\bfk,\bfk_{L}^{z}} - \delta_{\bfk,-\bfk_{L}^{z}}\right)
\right]  = 0   ~,\ee
where $\bfk_{L}^{0} = (k_{L},k_{L},k_{L})$,
 $\bfk_{L}^{x}= (k_{L},-k_{L},-k_{L})$, $\bfk_{L}^{y}= (-k_{L},k_{L},-k_{L})$ and $\bfk_{L}^{z}= (-k_{L},-k_{L},k_{L})$.
The qualitative behaviour of the solutions of \eq{aa2} is expected to be similar to the case of the planar inhomogeneity described by \eq{e13}, with the difference that each mode now directly couples to eight other modes rather than two. In particular, we expect that resonant behaviour will still occur for any mode $\bfk$ such that $|\bfk \pm \bfk_{L}^{i}|$ is small for $i = 0,\;x,\;y$ or $z$.

\renewcommand{\theequation}{B-\arabic{equation}}
 \setcounter{equation}{0}

\section*{Appendix B: Validity of the classical equation with space-dependence for quantum modes}

      In this Appendix we prove that the classical field equation \eq{e13a} correctly describes the
evolution of quantum modes in the presence of space-dependence.

        The action including the space-dependent step can be written as
$$    S = \frac{1}{2} \int d^{4}x \left[ u^{'\;2} - \left( \nabla u \right)^2 - m_{u}^{2} u^2 - m_{1}^{2}
\sin(\bfk_{L}.\bfx) u^2 \right. $$
 \be{a1} \left. - 2 a^{3} \frac{dF}{d \phi} {\rm sin(\bfk_{L}.\bfx)}u  \right]      ~,\ee
where $m_{u}^{2} = -z^{''}/z$ and $m_{1}^{2} = a^{2} d^{2}F/d \phi^{2}$.
The term linear in $u$ drives the growth of the modes with $\bfk = \pm \bfk_{L}$ and the modes
with $\bfk = n \bfk_{L}$ which are coupled to these. The growth of these modes, which is due to the classical evolution of the inflaton field in the space-dependent potential of the step, may be treated classically. We will therefore quantize the
action without the linear term, and consider the classical evolution of the modes with $\bfk = n\bfk_{L}$ separately.
 It will be convenient to use discrete modes in a box of volume $V$,
\be{a2} u(\bfx, \tau) = \frac{1}{\sqrt{V}} \sum_{\bfk} q(\bfk,\tau) e^{i\bfk.\bfx}     ~\ee
\be{a3} \pi(\bfx, \tau) = \frac{1}{\sqrt{V}} \sum_{\bfk} p(\bfk,\tau) e^{i\bfk.\bfx}     ~\ee
where $\pi = u^{'}$ is the conjugate momentum of $u$ and $q(-\bfk,\tau) = q^{\dagger}(\bfk,\tau)$.
The Hamiltonian for the modes in the absence of the linear term is then
$$ H = \frac{1}{2} \sum_{\bfk} \left[ p^{\dagger}\!(\bfk,\tau)p(\bfk,\tau)
+ (\bfk^{2} + m_{u}^{2}) q^{\dagger}\!(\bfk,\tau)q(\bfk,\tau)\right. $$
\be{a4a} \left. + \frac{m_{1}^{2}}{2 i}  q^{\dagger}\!(\bfk + \bfk_{L},\tau) q(\bfk,\tau)
- \frac{m_{1}^{2}}{2 i} q^{\dagger}\!(\bfk,\tau) q(\bfk + \bfk_{L} ,\tau)  \right]
~.\ee
This can be written in the form
\be{a4b} H = \frac{1}{2} \left[ P^{\dagger}P + Q^{\dagger}MQ \right]   ~\ee
where $P$ and $Q$ are column matrices with entries $p(\bfk,\tau)$ and $q(\bfk,\tau)$ labeled by $\bfk$ and $M$ is a Hermitian matrix.
The terms in \eq{a4a} proportional to $m_{1}^{2}$ are due to the space dependence. Since these terms mix modes with different $\bfk$, we cannot directly quantize the Hamiltonian by expanding $u(\bfx,\tau)$
in terms of creation and annihilation operators with mode functions $u_{\bfk}$ which satisfy the classical field equations.
Instead, we first transform $p$ and $q$ to new canonical coordinates and momenta which allow the Hamiltonian to be expressed as a sum over uncoupled quadratic Hamiltonians for each $\bfk$. To do this we make a unitary transformation of
$q$ and $p$ to new variables $s$ and $p_{s}$,
\be{a5}   S = UQ  \Rightarrow s(\bfk,\tau) = U_{\bfk,\bfk^{'}} q(\bfk^{'},\tau)   ~\ee
\be{a6}  P_{s} = U^{*}P \Rightarrow p_{s}(\bfk,\tau) = U^{*}_{\bfk,\bfk^{'}} p(\bfk^{'},\tau)    ~,\ee
where $U$ is a unitary matrix with rows and columns labeled by $\bfk$ which diagonalizes $M$.
The Hamiltonian can then be written in the form
\be{a8} H = \frac{1}{2} \sum_{\bfk} \left[ p_{s}^{\dagger}\!(\bfk,\tau)p_{s}(\bfk,\tau)
+ \tilde{\omega}_{\bfk}^{2} s^{\dagger}\!(\bfk,\tau)s(\bfk,\tau) \right] ~,\ee
where $\tilde{\omega}_{\bfk}^{2}(\bfk,\bfk_{L},\tau)$ are the eigenvalues of the matrix $M$. In addition,
$s$ and $p_{s}$ are conjugate variables, as can be seen from the
conjugate nature of $q$ and $p$,
$$ [p_{s}(\bfk,\tau), s(\bfl,\tau)] =
U^{*}_{\bfk,\bfk^{'}}  U_{\bfl,\bfl^{'}}  [p(\bfk^{'},\tau), q(\bfl^{'},\tau)] $$
$$ = U^{*}_{\bfk,\bfk^{'}}  U_{\bfl,\bfl^{'}}  \times -i \delta_{\bfk^{'},\bfl^{'}}
 = -i  U_{\bfl,\bfl^{'}} U^{\dagger}_{\bfl^{'},\bfk}  $$
$$  \Rightarrow  [p_{s}(\bfk,\tau), s(\bfl,\tau)] = -i \delta_{\bfl,\bfk}  $$
Since \eq{a8} is quadratic in the modes, we can quantize the system by expressing $p_{s}$ and $s$ in terms of annihilation operators and mode functions $s(\bfk,\tau) = \tilde{u}_{\bfk}(\tau) \tilde{a}_{\bfk}$, where $\tilde{u}_{\bfk}(\tau)$ will satisfy the
classical equation for $s(\bfk,\tau)$, obtained using the Hamiltonian form of the equations of motion for canonical coordinates and conjugate momenta, $p_{s}^{'} = -\partial H/\partial s$, $s^{'} = \partial H /\partial p_{s}$:
\be{a9} s^{''}\!\!(\bfk,\tau) = - \tilde{\omega}_{\bfk}^{2} s(\bfk,\tau)    ~.\ee
At times well before and after the step, $\tilde{u}_{\bfk}(\tau)$ and $u_{\bfk}(\tau)$ are the same, since then $s(\bfk,\tau) \rightarrow q(\bfk,\tau)$. Therefore the quantum initial conditions for $\tilde{u}_{\bfk}(\tau)$ are the same as those for $u_{\bfk}(\tau)$, given by \eq{e10}. By rotating back from S to Q via an inverse unitary transformation, the
classical equations \eq{a9} become the classical equations \eq{e13}. Therefore the solution of \eq{a9} for quantum initial conditions is the same as the solution of \eq{e13} with quantum initial conditions given by \eq{e10}. At times well after the step has occurred, $u_{\bfk}(\tau)$ may be interpreted as the quantum mode function with the field given by \eq{e7}, since then $s(\bfk,\tau) \rightarrow q(\bfk,\tau)$.

\renewcommand{\theequation}{C-\arabic{equation}}
 \setcounter{equation}{0}

\section*{Appendix C: Relating Modes to the Curvature Perturbation}

\subsection*{(i) Power Spectrum for Anisotropic Perturbations}

    The power spectrum of $u(\bfx,\tau)$ may be defined in terms of
the spatial average of $u(\bfx,\tau)^{2}$. Setting the initial phases to zero for the moment, from \eq{e7} this is given
by
\be{c1} \langle u(\bfx,\tau)^{2} \rangle \equiv \langle 0 \left| u(\bfx,\tau)^{2} \right| 0 \rangle = \int \!\!\frac{d^{3}k}{\left(2 \pi\right)^{3}} \left| u_{\bfk}
\right|^{2}        ~.\ee
In this we are applying the ergodic principle that classical spatial averages over a member of an ensemble are equal to quantum ensemble averages at fixed $\bfx$.
The mean value of $u(\bfx,\tau)^{2}$ is related to the power spectrum by
\be{c2}  \langle u(\bfx,\tau)^{2} \rangle = \int \!\!\frac{{\cal P}_{u}(\bfk)}{4 \pi \left| \bfk\right|^3} d^{3} k    ~.\ee
Since in most cases the power spectrum is independent of the direction of $\bfk$, ${\cal P}_{u}(\bfk)$ is usually
written as a function of $k = |\bfk|$. In the anisotropic case we can use the more general definition of \eq{c2}.
Comparing with \eq{c1} we then obtain
\be{c3}      {\cal P}_{u}(\bfk)   =  \frac{\left|\bfk\right|^{3}}{2 \pi^{2}}   \left| u_{\bfk}
\right|^{2}   ~.\ee
Therefore, since ${\cal R} = -u/z$, we find
\be{c4}  {\cal P_{R}}(\bfk) \equiv \frac{{\cal P}_{u}(\bfk)}{z^{2}} =  \frac{\left|\bfk\right|^{3}}{2 \pi^{2}}
\left| \frac{u_{\bfk}}{z}
\right|^2   ~.\ee

  In our calculation of the power spectrum, we apply \eq{c1} to the solution $u_{\bfk}$ corresponding to a particular set of random initial phases. The average of \eq{c1} is then an ensemble average of $u(\bfx,\tau)^{2}$ at fixed $\bfx$ over the classical universes generated by $u_{\bfk}$. Since the observed power spectrum is assumed to correspond to that obtained in one member of the ensemble by spatially averaging $u(\bfx,\tau)^{2}$, we are in practice applying the ergodic principle to the ensemble generated by each set of random phases i.e. we first fix the initial phases and then apply the ergodic principle to the resulting $u_{\bfk}(\tau)$.

\subsection*{(ii) Relating classical and quantum modes to the curvature perturbation}

  In the case of isotropic fluctuations, the power spectrum can directly interpreted as the contribution to $<{\cal R}^2>$ per logarithmic interval in $k$. However, we are considering here an anisotropic
step characterised by the direction $\bfk_{L}$.
In order to interpret $u_{\bfk}$ in this case, we can discretize $u(\bfx, \tau)$ into classical modes
\be{cl3}  u(\bfx,\tau) = \sum_{\bfk_{i}} u_{\bfk_{i}}^{cl}\, e^{i \bfk_{i}.\bfx}   ~.\ee
In this case
\be{cl4} \langle u(\bfx,\tau)^{2} \rangle  =   \sum_{\bfk_{i}} |u_{\bfk_{i}}^{cl}|^2      ~.\ee
For quantum modes we have
\be{cl5} \langle u(\bfx,\tau)^{2} \rangle \equiv \langle 0 \left| u(\bfx,\tau)^{2} \right| 0 \rangle = \int\!\! \frac{d^{3}k}{\left(2 \pi\right)^{3}} \left| u_{\bfk}
\right|^{2}        ~.\ee
To relate the quantum resonant modes to classical modes we can write this as a sum of the form \eq{cl4} with
\be{cl6} |u_{\bfk_{i}\;eff}^{cl}|^2 = \frac{1}{\left(2 \pi\right)^3}
\int_{-k_{L}/2}^{k_{L}/2} |u_{\bfk}|^2 d^{3} k_{i}^{'} \;\;\; ; \;\;\; \bfk_{i}^{'} = \bfk - \bfk_{i}       ~.\ee
This breaks the integral \eq{cl5} into cubes of side $k_{L}$ centered on points $\bfk_{i} = (a,b,c)k_{L}$ with $a,b,c$ integers.
In so far as the integral \eq{cl5} is dominated by the contributions of the resonances, we can then identify
$|u_{\bfk_{i}\;eff}^{cl}|$ in \eq{cl6} with the amplitude of effective classical modes in the sum \eq{cl3}
with $\bfk_{i} = n \bfk_{L}$.

   The resonances are a function of $|\bfk - n \bfk_{L}|$. Therefore
the integral over the volume of the cube \eq{cl6} can be approximated by an integral over a sphere of radius $k_{L}$.
With $|\ucl_{n\;eff}| \equiv |u_{n \bfk_{L}\;eff}^{cl}|$ and $\bfk^{'} \equiv \bfk - n \bfk_{L}$, \eq{cl6} becomes
\be{cl7} |\ucl_{n\;eff}|^2 = \frac{1}{\left(2 \pi\right)^3}
\int_{0}^{k_{L}} |u_{\bfk}|^2(k^{'}) \,\,4 \pi \,k^{'\;2} d k^{'}     ~.\ee
Since we can choose $\bfk^{'}$ along any direction, we can assume
$k^{'} = k_{x}-nk_{L}$ with $k_{x} > nk_{L}$.  
This allows us to directly compare the quantum modes with the
classical modes.

   The curvature perturbation due to the classical modes is then
\be{cl9}   {\cal R}(\bfx, \tau) = \sum_{\bfk_{n}} {\cal R}_{\bfk_{n}} e^{i \bfk_{n}.\bfx} \;\; ; \;\;\; \bfk_{n} = n \bfk_{L}
~,\ee
where
\be{cl10}  |{\cal R}_{\bfk_{n}}| = \frac{|\ucl_{n}|}{z}     ~.\ee
Therefore the planar inhomogeneity in the direction
$\bfk_{L}$ will result in a curvature perturbation which is a sum of modes of wavenumber $n \bfk_{L}$, which will manifest itself as anisotropic density perturbations in planes normal to the $\bfk_{L}$ direction.

    To constrain the amplitude of the classical modes $\ucl_{n}$,
we can interpret the observational constraint on $P_{{\cal R}}$
as a constraint on $<{\cal R}^2>$. For the case of
isotropic perturbations,
\be{cl11}  < {\cal R}^{2} >  = \int_{k_{min}}^{k_{max}} P_{{\cal R}} \frac{dk}{k}    ~,\ee
where $k_{min}$ and $k_{max}$ correspond to the largest and smallest scales observed in the CMB. Using the COBE normalized value,
$P_{{\cal R}}^{1/2}(k) = 4.8  \times 10^{-5}$ (essentially independent of $k$ for a near scale-invariant spectrum) and
$\ln(k_{max}/k_{min}) \approx 10$, we obtain
\be{cl12}  < {\cal R}^{2} >  = 10 P_{\cal R} = 2.3 \times 10^{-8}    ~.\ee
Since the contribution from the classical modes is
\be{cl13}  < {\cal R}^{2} > = \sum_{n} \frac{|\ucl_{n}|^{2}}{z^{2}}  ~\ee
we will therefore impose the conservative constraint
\be{cl14} \frac{|\ucl_{n}|}{z} \lae 10^{-4}    ~.\ee


\begin{thebibliography}{99}


\bibitem{adams}
J.~A.~Adams, B.~Cresswell and R.~Easther,
  Phys.\ Rev.\  D {\bf 64}, 123514 (2001)
  [arXiv:astro-ph/0102236].

\bibitem{peiris}
  H.~V.~Peiris {\it et al.}  [WMAP Collaboration],
  Astrophys.\ J.\ Suppl.\  {\bf 148} (2003) 213
  [arXiv:astro-ph/0302225].


\bibitem{step}
  L.~Covi, J.~Hamann, A.~Melchiorri, A.~Slosar and I.~Sorbera,
  Phys.\ Rev.\  D {\bf 74}, 083509 (2006)
  [arXiv:astro-ph/0606452].

\bibitem{step2}
  J.~Hamann, L.~Covi, A.~Melchiorri and A.~Slosar,
  Phys.\ Rev.\  D {\bf 76}, 023503 (2007)
  [arXiv:astro-ph/0701380].


\bibitem{joy}
  M.~Joy, V.~Sahni and A.~A.~Starobinsky,
  Phys.\ Rev.\  D {\bf 77} (2008) 023514
  [arXiv:0711.1585 [astro-ph]];
  M.~Joy, A.~Shafieloo, V.~Sahni and A.~A.~Starobinsky,
  arXiv:0807.3334 [astro-ph].


\bibitem{features}
 A.~A.~Starobinsky,
  JETP Lett.\  {\bf 55}, 489 (1992)
  [Pisma Zh.\ Eksp.\ Teor.\ Fiz.\  {\bf 55}, 477 (1992)];
 H.~M.~Hodges, G.~R.~Blumenthal, L.~A.~Kofman and J.~R.~Primack,
  Nucl.\ Phys.\  B {\bf 335}, 197 (1990);
  S.~M.~Leach and A.~R.~Liddle,
  Phys.\ Rev.\  D {\bf 63}, 043508 (2001)
  [arXiv:astro-ph/0010082];
 S.~M.~Leach, M.~Sasaki, D.~Wands and A.~R.~Liddle,
  Phys.\ Rev.\  D {\bf 64}, 023512 (2001)
  [arXiv:astro-ph/0101406];
 C.~P.~Burgess, J.~M.~Cline, F.~Lemieux and R.~Holman,
  JHEP {\bf 0302}, 048 (2003)
  [arXiv:hep-th/0210233];
  C.~R.~Contaldi, M.~Peloso, L.~Kofman and A.~Linde,
  JCAP {\bf 0307}, 002 (2003)
  [arXiv:astro-ph/0303636].

\bibitem{transient}
J.~M.~Cline, P.~Crotty and J.~Lesgourgues,
  JCAP {\bf 0309}, 010 (2003)
  [arXiv:astro-ph/0304558];
B.~A.~Powell and W.~H.~Kinney,
  Phys.\ Rev.\  D {\bf 76}, 063512 (2007)
  [arXiv:astro-ph/0612006].


\bibitem{tplanck}
  R.~H.~Brandenberger and J.~Martin,
  Mod.\ Phys.\ Lett.\  A {\bf 16}, 999 (2001)
  [arXiv:astro-ph/0005432];
  J.~Martin and R.~H.~Brandenberger,
  Phys.\ Rev.\  D {\bf 63}, 123501 (2001)
  [arXiv:hep-th/0005209];
  R.~Easther, B.~R.~Greene, W.~H.~Kinney and G.~Shiu,
  Phys.\ Rev.\  D {\bf 64}, 103502 (2001)
  [arXiv:hep-th/0104102];
  Phys.\ Rev.\  D {\bf 66}, 023518 (2002)
  [arXiv:hep-th/0204129];
  U.~H.~Danielsson,
  Phys.\ Rev.\  D {\bf 66}, 023511 (2002)
  [arXiv:hep-th/0203198].





\bibitem{hi}
J.~A.~Adams, G.~G.~Ross and S.~Sarkar,
  Nucl.\ Phys.\  B {\bf 503}, 405 (1997)
  [arXiv:hep-ph/9704286];
 J.~Lesgourgues,
  Nucl.\ Phys.\  B {\bf 582}, 593 (2000)
  [arXiv:hep-ph/9911447].

\bibitem{hunt}
 P.~Hunt and S.~Sarkar,
  Phys.\ Rev.\  D {\bf 70}, 103518 (2004)
  [arXiv:astro-ph/0408138];
  Phys.\ Rev.\  D {\bf 76} (2007) 123504
  [arXiv:0706.2443 [astro-ph]].





\bibitem{tp} G.~N.~Felder, J.~Garcia-Bellido, P.~B.~Greene, L.~Kofman, A.~D.~Linde and I.~Tkachev,
  Phys.\ Rev.\ Lett.\  {\bf 87}, 011601 (2001)
  [arXiv:hep-ph/0012142];
  G.~N.~Felder, L.~Kofman and A.~D.~Linde,
  Phys.\ Rev.\  D {\bf 64}, 123517 (2001)
  [arXiv:hep-th/0106179].



\bibitem{hitp}
  M.~Broadhead and J.~McDonald,
  Phys.\ Rev.\  D {\bf 68}, 083502 (2003)
  [arXiv:hep-ph/0305298];
  Phys.\ Rev.\  D {\bf 72} (2005) 043519
  [arXiv:hep-ph/0503081].



\bibitem{muk}
 V.~F.~Mukhanov,
  Phys.\ Lett.\  B {\bf 218} (1989) 17;
  V.~F.~Mukhanov, H.~A.~Feldman and R.~H.~Brandenberger,
  Phys.\ Rept.\  {\bf 215}, 203 (1992).


\bibitem{pol} D.~Polarski and A.~A.~Starobinsky,
  Class.\ Quant.\ Grav.\  {\bf 13}, 377 (1996)
  [arXiv:gr-qc/9504030].


\bibitem{stepNG1}
X.~Chen, R.~Easther and E.~A.~Lim,
  JCAP {\bf 0706}, 023 (2007)
  [arXiv:astro-ph/0611645].

\bibitem{stepNG2} X.~Chen, R.~Easther and E.~A.~Lim,
  JCAP {\bf 0804}, 010 (2008)
  [arXiv:0801.3295 [astro-ph]].


\bibitem{asym}
H.~K.~Eriksen, F.~K.~Hansen, A.~J.~Banday, K.~M.~Gorski and P.~B.~Lilje,
  Astrophys.\ J.\  {\bf 605} (2004) 14
  [Erratum-ibid.\  {\bf 609} (2004) 1198]
  [arXiv:astro-ph/0307507];
 F.~K.~Hansen, A.~J.~Banday and K.~M.~Gorski,
  arXiv:astro-ph/0404206;
H.~K.~Eriksen, A.~J.~Banday, K.~M.~Gorski, F.~K.~Hansen and P.~B.~Lilje,
  Astrophys.\ J.\  {\bf 660}, L81 (2007)
  [arXiv:astro-ph/0701089].



\bibitem{kam}
 A.~L.~Erickcek, M.~Kamionkowski and S.~M.~Carroll,
  arXiv:0806.0377 [astro-ph];
 A.~L.~Erickcek, S.~M.~Carroll and M.~Kamionkowski,
  arXiv:0808.1570 [astro-ph].

\bibitem{matsuda}
T.~Matsuda,
arXiv:0811.1318 [hep-ph].


\bibitem{poltest}
  C.~Dvorkin, H.~V.~Peiris and W.~Hu,
  Phys.\ Rev.\  D {\bf 77} (2008) 063008
  [arXiv:0711.2321 [astro-ph]].










\end{thebibliography}
\end{document}